\documentclass[preprint,12pt]{elsarticle}

\usepackage{amssymb}

\usepackage{amsmath}
\usepackage{lmodern}  
\usepackage{color}

\journal{International Journal of Heat and Mass Transfer}

\newcommand\nc{\newcommand}
\nc\includegraphicss{\includegraphics}
\nc\qlam{k}
\nc\qtau{\tau_q}

\begin{document}

\begin{frontmatter}



\title{Implicit numerical schemes for generalized heat conduction equations}


\author{\'A.~Rieth$^{1}$, R. Kov\'acs$^{123}$, T. F\"ul\"op$^{13}$}

\address{%
$^1$Department of Energy Engineering, Faculty of Mechanical Engineering, BME, Budapest, Hungary\\
$^2$Department of Theoretical Physics, Institute for Particle and Nuclear Physics, Wigner Research Centre for Physics, Budapest, Hungary \\
$^3$Montavid Thermodynamic Research Group, Budapest, Hungary}

 \begin{abstract}
There are various situations where the classical Fourier's law for heat
conduction is not applicable, such as heat conduction in heterogeneous
materials \cite{Botetal16, Vanetal17} or for modeling low-temperature
phenomena \cite{KovVan15, KovVan16, KovVan18}. In such cases, heat flux
is not
 directly
proportional to temperature gradient, hence,
the role -- and both the analytical and numerical treatment -- of
boundary conditions becomes nontrivial. Here, we address this question
for finite difference numerics via a shifted field approach. Based on
this ground,
implicit schemes are presented and compared to each other for the
Guyer--Krumhansl generalized heat conduction equation, which
successfully describes numerous beyond-Fourier experimental findings.
The results are validated by an analytical solution, and are contrasted
to finite element method
 outcomes
obtained by COMSOL. 
 \end{abstract}

\begin{keyword}
Implicit scheme \sep shifted fields \sep boundary conditions \sep nonequilibrium thermodynamics


\end{keyword}

\end{frontmatter}


\section{Introduction}
\label{intro}

The need to go beyond the Fourier heat conduction equation -- which reads
in one spatial dimension
 \begin{equation}  \label{foueq}
\partial_t T = \alpha \partial_{xx } T
 \end{equation}
for temperature $T$, with thermal diffusivity $\alpha$, and which
contains only first order time derivative $\partial_t$ and second order
space derivative $\partial_{xx}$ --
is experimentally proved under various conditions since decades
\cite{Botetal16, Vanetal17, Pesh44,JacWalMcN70, JacWal71, NarDyn72a,
Kam90, Jaetal08}. These circumstances are related partly to the material
structure \cite{Mari14, PMMar17conf}, and partly to the environment like
temperature and excitation \cite{myphd2017}. The characteristics of the
interaction between the sample and the environment are condensed into
the boundary conditions,
the role of which are
 therefore
crucial during the modeling.

For theories beyond the Fourier one, the common starting point is the
balance equation of internal energy $e$,
 \begin{equation}  \label{baleneq}
\rho \partial_t e + \partial_x q = 0,
 \end{equation}
also written in one spatial dimension, with density $\rho$ and heat flux
$q$. For many applications, a constant specific heat $c$ can be assumed,
yielding $e=cT$.

Then,
if one takes Fourier's law,
 \begin{equation}  \label{foueq2}
q = - \qlam \partial_x T,
 \end{equation}
where $\qlam$ is thermal conductivity, then (\ref{foueq}) can be
obtained.
In parallel, heat flux boundary conditions -- like a heat
pulse on one end and an adiabatic insulation on the other one, the case
considered hereafter -- can be written directly for temperature,
prescribing its gradient.

However, for generalized heat conduction models,
 the picture is not so simple
any more.
 For example, in
the first known extension to Fourier's law,
the so-called
Maxwell--Cattaneo--Vernotte (MCV) constitutive equation \cite{Max1867,
Cattaneo58, Vernotte58, Gyar77a}
 \begin{equation}  \label{mcveq}
\qtau \partial_t q + q = - \qlam \partial_x T,
 \end{equation}
 time derivative of heat flux also
appears, accompanied by a coefficient $\qtau$ called relaxation time.
In this case a heat flux boundary condition cannot be translated to
a Neumann-type boundary condition on temperature.
The situation becomes even more involved for the
Guyer--Krumhansl (GK) equation \cite{GuyKru66a1, GuyKru66a2, Van01a},
 \begin{equation}  \label{gkeq}
\qtau \partial_t q + q = - \qlam \partial_x T + \kappa^2 \partial_{xx} q,
 \end{equation}
where $\kappa^2$ is a parameter strongly related to the mean free path
from the aspect of kinetic theory \cite{MulRug98}. According to
room-temperature experiments \cite{Botetal16, Vanetal17, myphd2017},
measured deviation from the Fourier prediction always occurs in the
overdamped ($\kappa^2>\qtau$) region (as opposed to
the near-to-MCV region $\kappa^2<\qtau$), thus
usage of the GK equation is inescapable.

\begin{figure}[t]
\centering
\includegraphics[width=.75\textwidth]{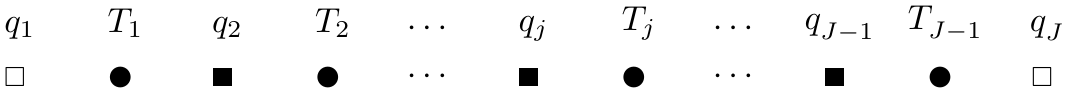}
\caption{Discretization method of shifted fields for a heat pulse
setting. Prescribed boundary values are $q_1$ as a function
of time, and $q_J = 0$ representing adiabatic insulation. All other
values (illustrated by filled rectangles and circles) can be computed
from neighboring and previous values. Temperature values sit at cell
midpoints
 while
heat flux values reside at cell boundaries.}
\label{fig:xxx}
\end{figure}

Combining (\ref{gkeq}) with (\ref{baleneq}) provides the
temperature-only version of the GK model:
 \begin{equation}  \label{gkeq2}
\qtau \partial_{tt} T + \partial_t T = \alpha \partial_{xx} T + \kappa^2 \partial_{txx} T.
 \end{equation}
Solving this equation
with
heat flux boundary conditions, especially with time dependent ones
needed for evaluating heat pulse experiments \cite{Botetal16,
Vanetal17}, is difficult. It is not clear how to translate conditions on
$q(t)$ to temperature $T(t)$, the two quantities being related to one
another according to a constitutive equation (\ref{gkeq}).
This was the motivation to develop a simple and fast numerical scheme, a
scheme of shifted fields \cite{KovVan15}, that was specifically devised
to be suitable for this type of problem.
The term shifted fields
refers here to
the spatial discretization method. Namely, instead of solving (\ref{gkeq2})
for $T$, the set of equations (\ref{baleneq}) and (\ref{gkeq}) are solved
for $T$ and $q$ both,
where
spatial locations of 
temperature values are shifted by a half space step with respect to
locations of $q$ values (see Fig.~\ref{fig:xxx}). This enables
us to prescribe boundary conditions only for
heat flux.

A physical interpretation of such a scheme is the distinction between
surface-related and volume-related quantities of the discrete cells.
More closely, 
temperature represents the average value over the volume while heat flux
describes energy flow at the cell boundary.

Notably,
similar but different schemes
like the two-step Lax-Wendroff, leapfrog or
Finite-Difference-Time-Domain (FDTD) methods are known in the
literature. All these apply values at half time step or half space step
to update a grid point at the next time instant \cite{NumRec07b}.
Furthermore, the present shifted field concept also differs from the
multigrid method where the goal is to increase accuracy by applying
finer and finer meshes \cite{NumRec07b, Chapra98}. One should also
mention Feynman \cite{Feynman1b} who presents a technique for time
integration for the dynamics of a point particle where the shifting is
used for time steps only. Moreover, Yee discusses the problem of
electromagnetic wave propagation and applies the FDTD method to solve
the Maxwell equations, and also discusses the possible boundary
conditions for such wave propagation problem \cite{Yee97}.
 However, none of the mentioned techniques address
the question of boundary conditions, and the advantage of the shifted
strategy for boundary conditions -- especially such nontrivial ones --
is not realized.
One should also pay attention to the work of Berezovski et al. \cite{KolEta17a, BerEta10p, Ber11a, BerVan17p, BerVan17b} where remarkably efficient numerical schemes are developed and tested for wave propagation problems. 

Our approach uses simple finite differences to approximate the partial
derivatives. An explicit version has already been
developed \cite{KovVan15}; here we present the realization of the
corresponding implicit version, which turns out to be remarkably superior
in performance aspects.
The outcomes are also compared to analytical
and finite element solutions in respect of efficiency and speed.

\section{Explicit scheme}

For the explicit scheme,
all related analysis and detailed discussion are published in
\cite{KovVan15},
and are only summarized
here for the sake of completeness. Hereafter,
dimensionless quantities \cite{KovVan15} are
used, which is a framework that is simple yet satisfactory for the
current numerics-related considerations.

The discretized form of the balance equation of internal energy
(\ref{baleneq}) is
 \begin{equation}  \label{disc:inten}
T^{n+1}_j=T^n_j-\frac{\Delta t}{\tau_{\Delta} \Delta x} ( q^n_{j+1}-q^n_j),
 \end{equation}
where $n$ indexes
time steps and $j$
 the space steps,
$\tau_\Delta$
denotes the dimensionless pulse
duration time
and $\Delta t$ the time step.
The Guyer--Krumhansl
constitutive equation is discretized as
 \begin{equation}  \label{GKdisc}
q^{n+1}_j=q^n_j-\frac{\Delta t}{\qtau} q^n_j - \frac{\tau_{\Delta}
\Delta t}{\qtau \Delta x} (T^n_j -T^n_{j-1}) + \frac{\kappa^2 \Delta
t}{\qtau \Delta x^2} (q^n_{j+1}-2q^n_j +q^n_{j-1}) ,
 \end{equation}
which is able to reproduce the solutions of the MCV model ($\kappa=0$)
and of the Fourier one ($\qtau = \kappa^2$). In these formulae, forward
time differencing is applied, which makes all the schemes first order in
time. Let us draw attention again to the boundary conditions, thanks to
which $T$ values can be updated without prescribing anything for
temperature at the boundaries. 

The scheme being explicit, one has to calculate the stability criteria
as well.
In \cite{KovVan15}, von
Neumann and Jury methods \cite{NumRec07b, Jury74} are used to determine
the stability conditions. In order to prove the convergence of such a
scheme, the Lax--Richtmyer theorem \cite{Lax56stab} is exploited by
proving the consistency of the schemes together with their stability.
Regarding consistency, although only its weak form is proved
\cite{myphd2017}, it is enough to fulfill the Lax--Richtmyer theorem and
ensure the presence of convergence \cite{FDA1, Gerdt12}.

\section{Implicit schemes}

When
quantities at
time instant $t^{n+1}$ are also considered, the scheme becomes implicit,
leading to the following discretized form of the balance equation of
internal energy:
 \begin{align}  \label{IMPSEPEN}
\tau_{\Delta} \frac{1}{\Delta t} (T^{n+1}_j - T^n_j) = -\frac{1}{\Delta
x} \left [ (1-\Theta) \left( q^n_{j+1} - q^n_j \right) +
\Theta \left( q^{n+1}_{j+1} -
q^{n+1}_j \right) \right ] ,
 \end{align}
and of the GK-type constitutive equation:
 \begin{align}  \label{IMPSEPGK}
\frac{\qtau}{\Delta t} \left ( q^{n+1}_j - q^n_j \right )
+ \left [ (1-\Theta) q^n_j+ \Theta q^{n+1}_j \right ] &
 \nonumber \\
+ \frac{\tau_{\Delta}}{\Delta x} \left [ (1-\Theta) ( T^n_j - T^n_{j-1}) +
\Theta (T^{n+1}_j- T^{n+1}_{j-1}) \right ] &
 \\ \nonumber
- \frac{\kappa^2}{\Delta x^2} \left [ (1-\Theta) \big ( q^n_{j+1} - 2q^n_j + q^n_{j-1} \big ) + \Theta \big (q^{n+1}_{j+1} - 2q^{n+1}_j + q^{n+1}_{j-1} \big ) \right ]
& = 0,
 \end{align}
where the convex combination of explicit and implicit terms is
characterized by the parameter $\Theta$, with $\Theta=0$
removing
the implicit terms and returning the purely explicit scheme.
Analogously, for $\Theta=1$, all the explicit terms vanish, making
(\ref{IMPSEPEN}) and (\ref{IMPSEPGK}) purely implicit. Choosing $\Theta
= 1/2$ gives the so-called
Crank--Nicolson scheme, which preserves the unconditionally stable
property of implicit schemes and provides
one order higher accuracy.
Here, accuracy is not analyzed in detail. We test the implicit scheme
with settings $\Theta=1/2$ and $\Theta = 1$, for various parameter
values for $\qtau$ and $\kappa^2$. 

In order to prove that no stability condition is needed for these
implicit schemes, we use the methods of von Neumann and Jury as before
\cite{KovVan15}, i.e., let us assume the solution of the difference
equations (\ref{IMPSEPEN}) and (\ref{IMPSEPGK}) in the form 
 \begin{equation}  \label{NSOL}
\phi^n_j=\xi^n e^{ikj \Delta x},
 \end{equation}
where $i$ is the imaginary unit, $k$ is the wave number parameter of the
solution, $j \Delta x$ denotes the $j^{\text{th}}$ discrete spatial
position,
and the complex number $\xi$ is called the growth factor
\cite{NumRec07b}. The scheme is stable if and only if $|\xi| \leq 1$
holds.
Now, using
(\ref{NSOL}) one can express each term from
(\ref{IMPSEPEN})--(\ref{IMPSEPGK}), for example $q^{n+1}_{j+1} =
\xi^{n+1} e^{ik(j+1)\Delta x} \cdot q_0$. Substituting back
(\ref{NSOL}) into (\ref{IMPSEPEN}) and (\ref{IMPSEPGK}) yields
 \begin{align}  \label{DISCSYS}
T_0 (\xi -1) + q_0 \frac{\Delta t}{\tau_{\Delta} \Delta x} \left [ ( 1-
\Theta ) \big ( e^{i k \Delta x} -1 \big ) + \Theta \xi \big ( e^{i k
\Delta x} - 1 \big ) \right ] & = 0,
 \\
q_0 ( \xi -1) + q_0 \frac{\Delta t}{\qtau}
\left [ (1-\Theta) + \Theta \xi \right ] &
 \nonumber \\
+ T_0 \frac{\tau_{\Delta} \Delta t}{\Delta x \qtau} \left [ (1-\Theta)
\big ( 1- e^{-i k \Delta x} \big ) + \Theta \xi \big (1- e^{-i k \Delta
x} \big ) \right ]
 \nonumber \\
- q_0 \frac{\kappa^2 \Delta t}{\Delta x^2 \qtau} \left [ ( 1-\Theta)
\left ( e^{i k \Delta x} -2 + e^{- i k \Delta x} \right ) + \Theta \xi
\left (e^{i k \Delta x} - 2 + e^{-i k \Delta x} \right ) \right ] & = 0.
 \end{align}
Then constructing a coefficient matrix and calculating its determinant
leads to the characteristic polynomial of system (\ref{DISCSYS}) in the
form $F(\xi) = a_2 \xi^2 + a_1 \xi + a_0$ with
 the coefficients:
coefficients
 \begin{align}
a_0 & = 1- \frac{\Delta t}{\qtau} ( 1-\Theta) +\left [ 2 \cos(k \Delta
x) -2 \right ] \left ( 1 - \Theta \right ) \frac{\Delta t}{\Delta x^2
\qtau} \left [ \kappa^2 - \Delta t ( 1-\Theta) \right ],
 \nonumber \\
a_1 & = -2 + \frac{\Delta t}{\qtau} (1- 2 \Theta) +
\left [ 2 \cos(k \Delta x) -2 \right ]
\frac{\Delta t}{\Delta x^2 \qtau} \left [ \kappa^2 (2 \Theta -1)
- 2 \Delta t \left ( 1 - \Theta \right ) \Theta \right ] ,
 \nonumber \\
a_2 & = 1 + \frac{\Delta t}{\qtau} \Theta - \left [
2 \cos(k \Delta x) -2 \right ]  \frac{\Delta t}{\Delta x^2 \qtau} \Theta \big (\kappa^2 +\Delta t \Theta \big ).
 \end{align}
The Jury criteria \cite{Jury74} can be used to obtain the requirements
in order to ensure that the roots of characteristic polynomial remain
within the unit circle in the complex plane. These criteria are, for the
polynomial $F$:
 \begin{enumerate}
\item $F(\xi=1) > 0$,
\item $F(\xi=-1) > 0$,
\item $|a_0| < a_2$.
 \end{enumerate}
Calculating each condition for $\Theta=1$ gives us 
 \begin{enumerate}
\item $\frac{4 \Delta t^2}{\qtau \Delta x^2}>0$, 
\item $4 + 2\frac{\Delta t}{\qtau} + 4 \frac{\Delta t}{\qtau \Delta x^2} ( 2\kappa^2 + \Delta t) >0$, 
\item $1 < 1 + \frac{\Delta t}{\qtau} + 4 \frac{\Delta t}{\qtau \Delta x^2} ( \kappa^2 + \Delta t)$, 
 \end{enumerate}
hence, the scheme has met the requirements
 as long as all parameters are positive.
In case of $\Theta=1/2$, we have
 \begin{enumerate}
\item $\frac{4 \Delta t^2}{\qtau \Delta x^2}>0$,
\item $4 >0$,
\item  $0 < 1 + \frac{4 \kappa^2}{\Delta x^2}$, 
 \end{enumerate}
that is, the first Jury criterion gives the same result and the other
two conditions are simpler and naturally fulfilled again.
We remark that,
 for the MCV equation
($\kappa=0$), each criteria are fulfilled, too. Therefore, the schemes are
stable and convergent.

\section{Comparison with analytical solutions}

Analytical solution for the GK equation is known for several cases
\cite{Zhukov16, Zhu16a, Zhu16b, ZhuSri17}, even for boundary conditions
related to heat pulse experiments with adiabatic condition on the rear
side \cite{Kov18gk}.
The analytical solution is available in an infinite sum form
\cite{Kov18gk}.
 For the benchmark comparison between analytical and numerical solutions
presented here, the following parameters have been applied:
 \begin{enumerate}
\item Solution of Fourier equation: $\qtau = \kappa^2$, 
\item Solution of MCV equation: $\qtau=0.02$, $\kappa^2=0$,
\item Over-diffusive solution of the GK equation: $\qtau=0.02$,
$\kappa^2=10 \qtau$.
 \end{enumerate}
Moreover, heat pulse duration $\tau_\Delta=0.04$ is used in all cases,
and the simulated time interval (dimensionless time $t =1$) and the
number of cells ($300$) are also fixed. The various schemes are
compared based on the computational run time measured by MATLAB.
Although the run time itself is not representative in a single scheme
and depends on many other conditions like programming language,
realization of a scheme, properties of hardware, etc., for comparative
reasons it is useful and representative.  It is important to emphasize
that the over-diffusive range ($\kappa^2>\qtau$) is
distinguished by experiments, i.e., the measured non-Fourier behavior
always occurs in that region of parameters. 

\begin{figure}
\centering
\includegraphicss[width=12cm,height=7cm]{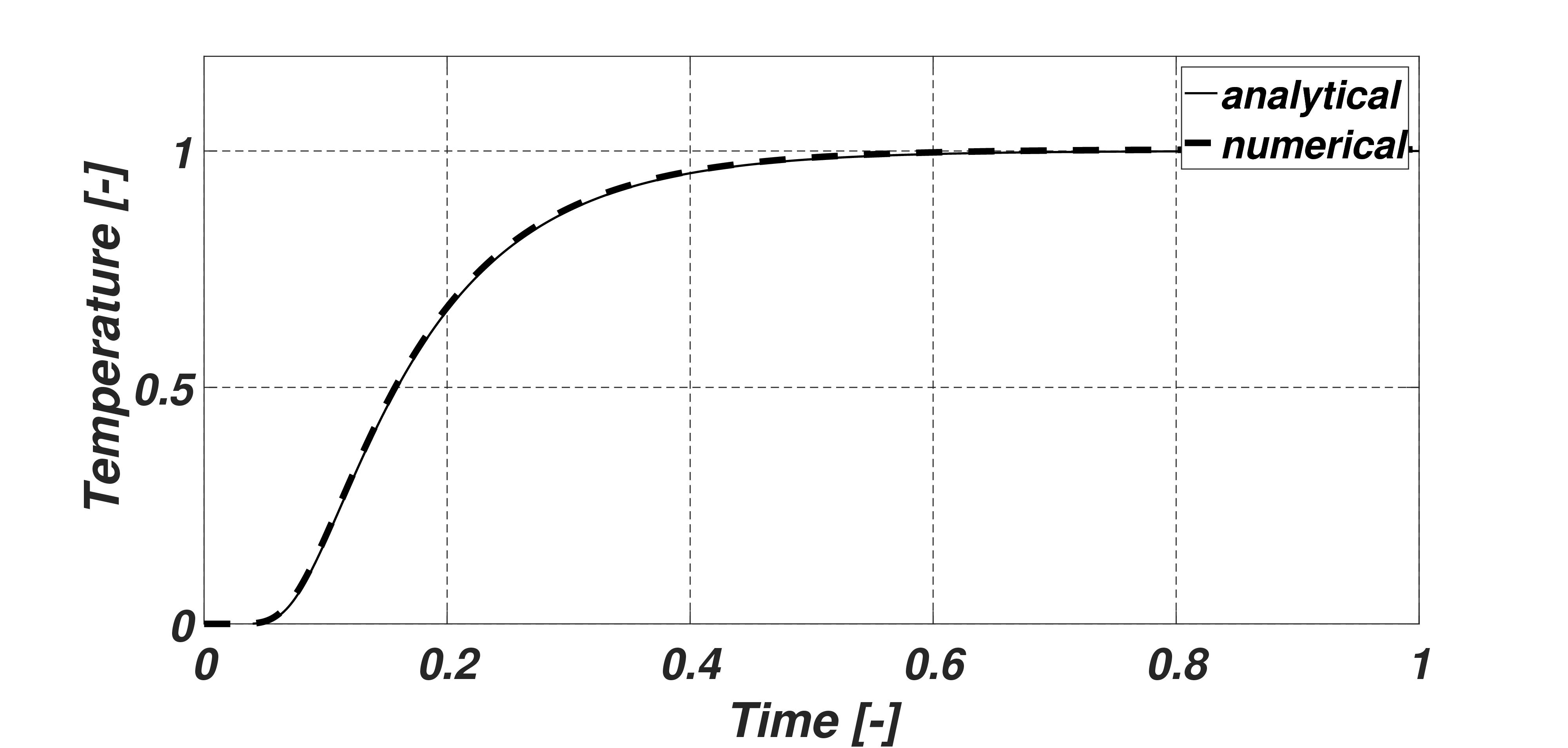}
\caption{Rear-side temperature histories:
the analytical solution (solid line) and the $\Theta=1$ scheme (thicker
dashed line).}
 \label{fig:impfou_anal}
\end{figure}

\begin{figure}
\centering
\includegraphicss[width=12cm,height=7cm]{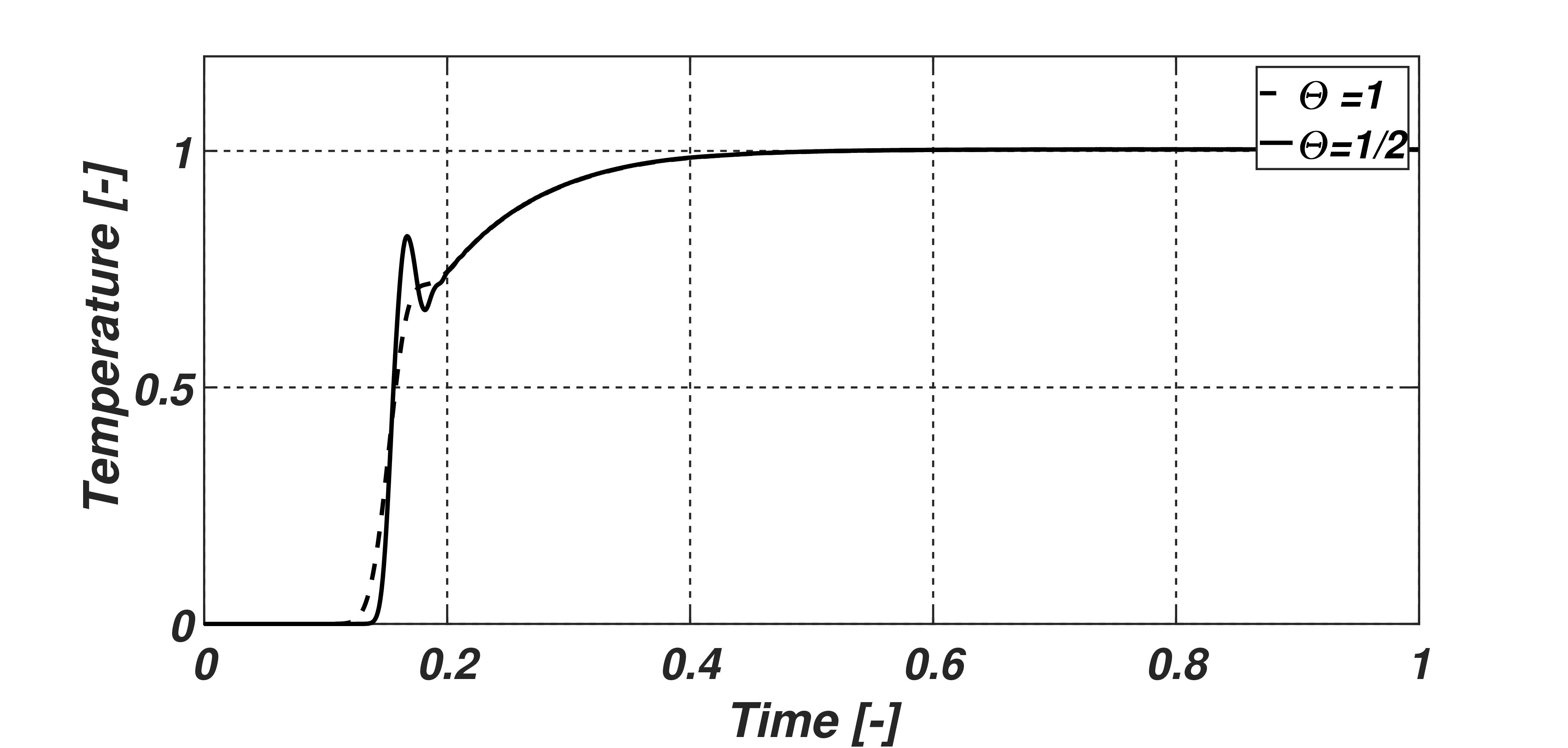}
\caption{Rear-side temperature histories when solving the MCV equation
with schemes $\Theta=1$ ($1000$ time steps) and $\Theta=1/2$. The dashed line belongs to $\Theta=1$.}
\label{fig:impmcv1}
\end{figure}

\begin{figure}
\centering
\includegraphicss[width=12cm,height=7cm]{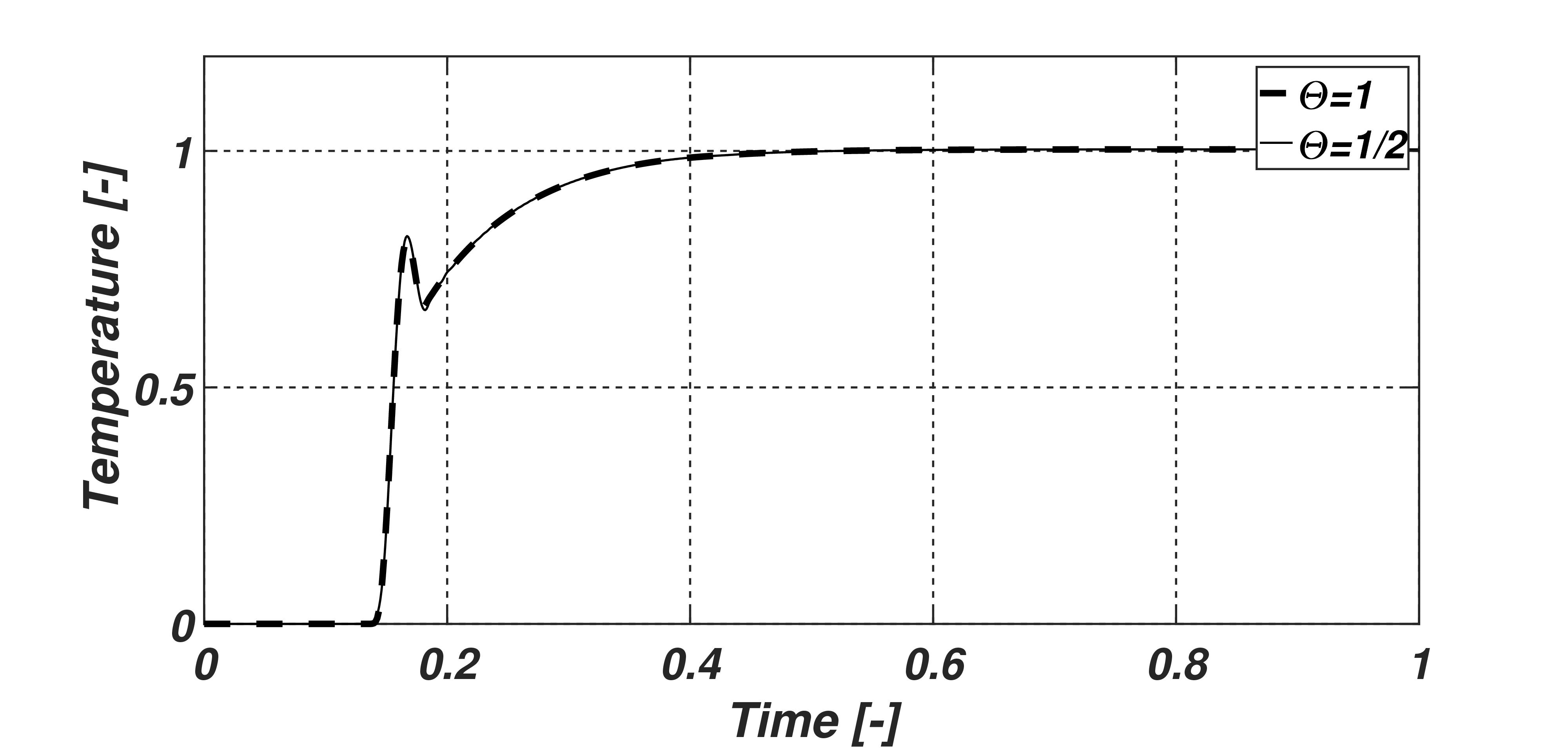}
\caption{Rear-side temperature histories when solving the MCV equation
with schemes $\Theta=1$ ($10^5$ time steps) and $\Theta=1/2$. The dashed line belongs to $\Theta=1$.}
\label{fig:impmcv2}
\end{figure}

\begin{figure}
\centering
\includegraphicss[width=12cm,height=7cm]{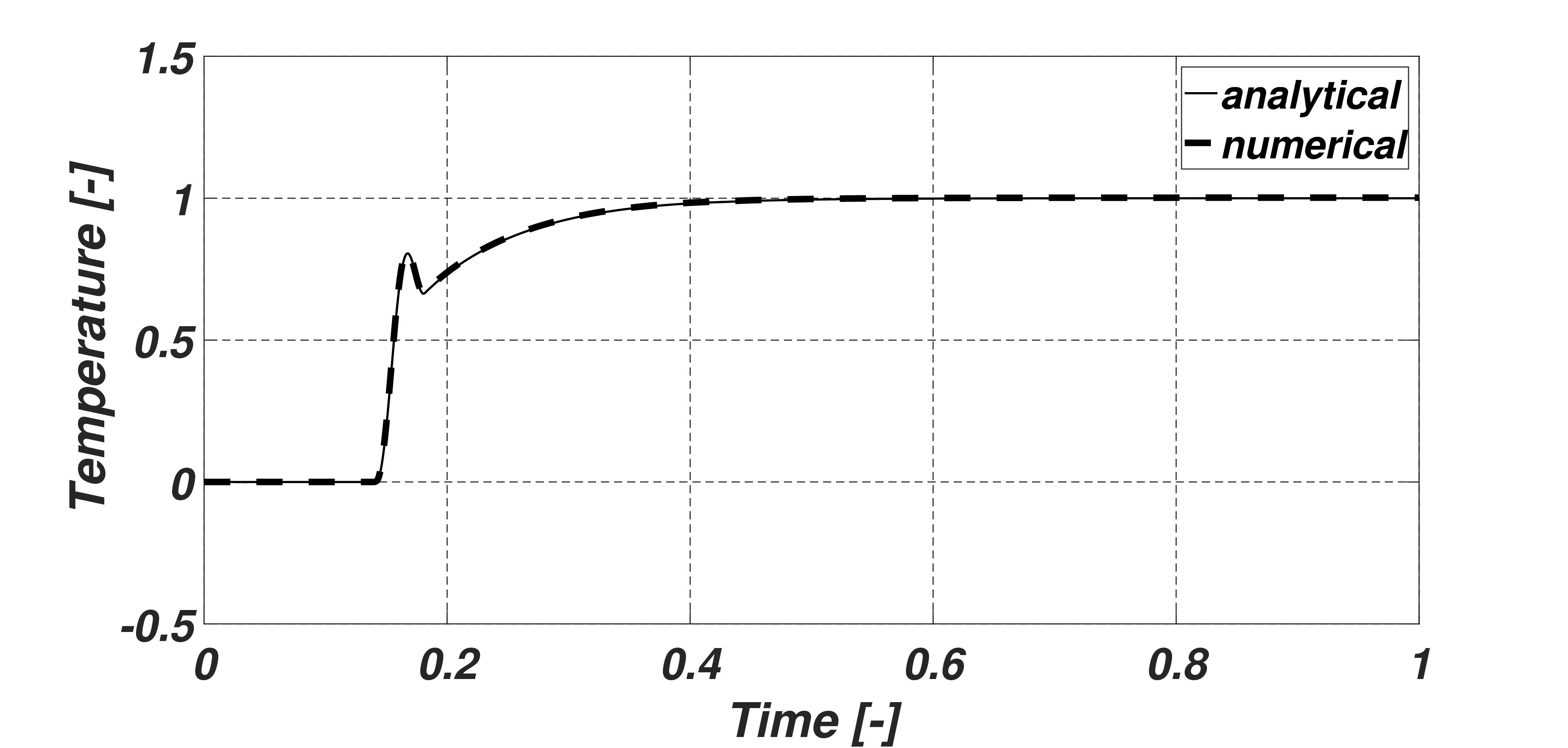}
\caption{Rear-side temperature histories: comparison between the
analytical solution (solid line) and the $\Theta=1/2$ scheme (thicker dashed line).}
\label{fig:impmcv_anal}
\end{figure}

\begin{figure}
\centering
\includegraphicss[width=12cm,height=7cm]{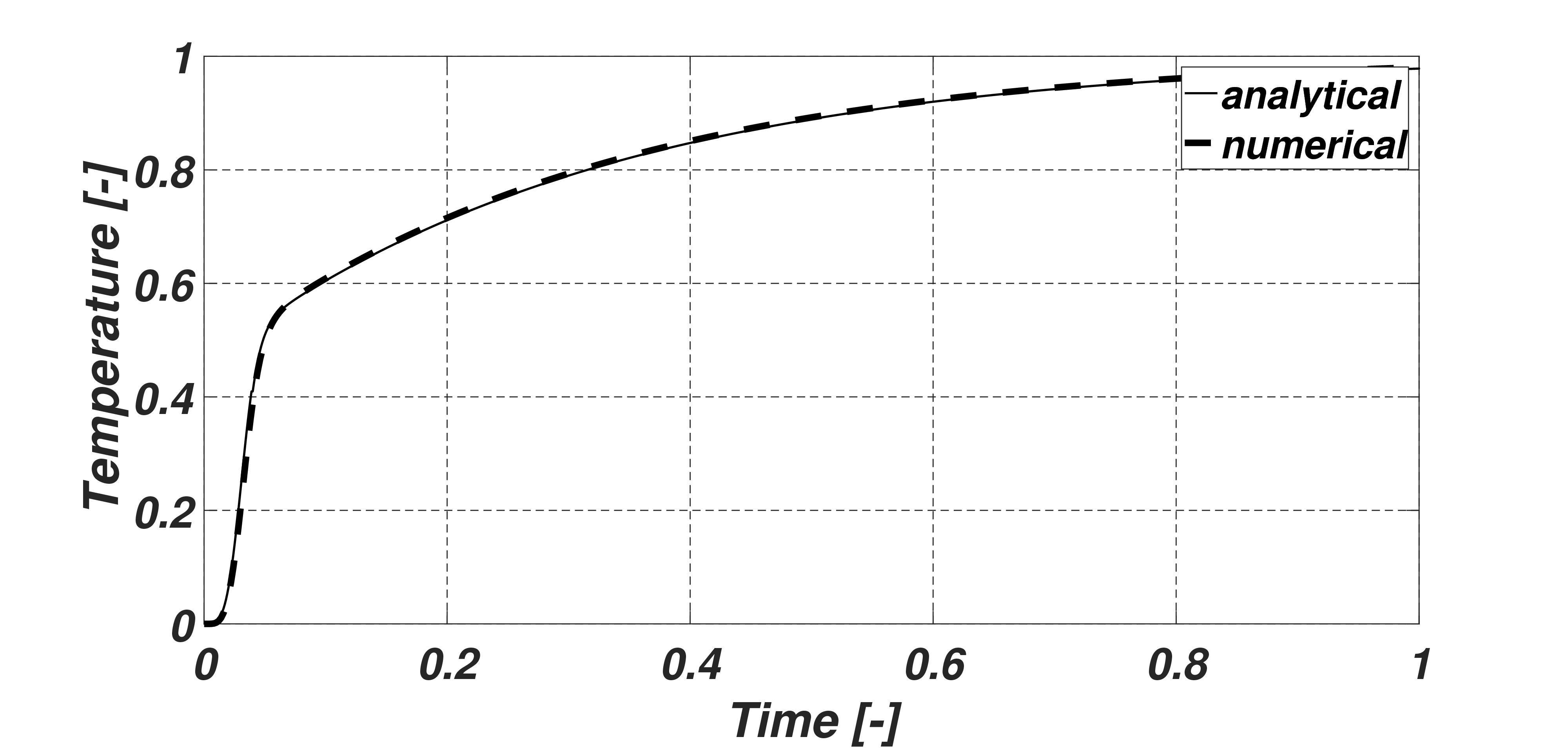}
\caption{Rear-side temperature histories: comparison between the
analytical solution (solid line) and the $\Theta=1$ scheme (thicker dashed line).}
\label{fig:impgk_anal}
\end{figure}

 \begin{enumerate}
\item Case of the Fourier equation:
 \begin{enumerate}
\item $\Theta=1$ scheme requires $100$ time steps and the solution takes $0.119$ s (see
Fig.~\ref{fig:impfou_anal}). 
\item $\Theta=1/2$ scheme shows no difference either in accuracy or in run time. 
\item $\Theta=0$ explicit scheme requires ca.\ $10^6$ time steps, which takes $142.9$ s. 
\item The analytical solution requires $50$ terms and takes $0.08$ s with $100$ time steps (see Fig.~\ref{fig:impfou_anal}).
 \end{enumerate}
\item Case of the MCV equation:
 \begin{enumerate}
\item For $\Theta=1$ scheme, $1000$ time steps are not sufficient.
The solution was not accurate enough (see Fig.~\ref{fig:impmcv1}).
With a new setting of $10^5$ time steps, solution takes $15.4$ s (Fig.~\ref{fig:impmcv2}).
\item $\Theta=1/2$ scheme shows significant difference especially for
hyperbolic equations like the MCV one. The vicinity of the wave front
is more accurate than in the previous case, $1000$ time steps are sufficient and it requires $0.2$ s.
\item $\Theta=0$ explicit scheme requires ca.\ $10^6$ time steps again, which takes $145.5$ s. 
\item The analytical solution requires $200$ terms and takes $6.6$ s with $500$ time steps (see Fig.~\ref{fig:impmcv_anal}).
 \end{enumerate}
\item Case of the GK equation:
 \begin{enumerate}
\item $\Theta=1$ scheme requires $100$ time steps and the solution takes $0.261$ s (see Fig.~\ref{fig:impgk_anal}). 
\item $\Theta=1/2$ scheme shows no difference either in accuracy or in run time. 
\item $\Theta=0$ explicit scheme requires ca.\ $10^7$ time steps, which takes $1640$ s. 
\item The analytical solution requires $5$ terms and takes $0.07$ s with $200$ time steps (see Fig.~\ref{fig:impgk_anal}).
\end{enumerate}
\end{enumerate}


As it is clear, the implicit schemes reproduce the analytical solution
in every case. In fact, they could be faster for solutions containing
jumps like in case of the MCV equation. Moreover, the capabilities of
the analytical solution approach are more limited -- for example, the GK equation for finite time
heat pulse excitation with cooling boundary conditions is not yet
solved. In such cases the numerical methods are the only way to obtain
the solution. It is important to observe the significant difference
between $\Theta=1$ and $1/2$ schemes for hyperbolic equations. The
explicit scheme was the slowest and less efficient, not surprisingly.

\section{Comparison with finite element method}

In this section, the finite element implementation of
the same problem is presented, using the software COMSOL v5.3a.

\begin{figure}
\centering
\includegraphicss[width=12cm,height=7cm]{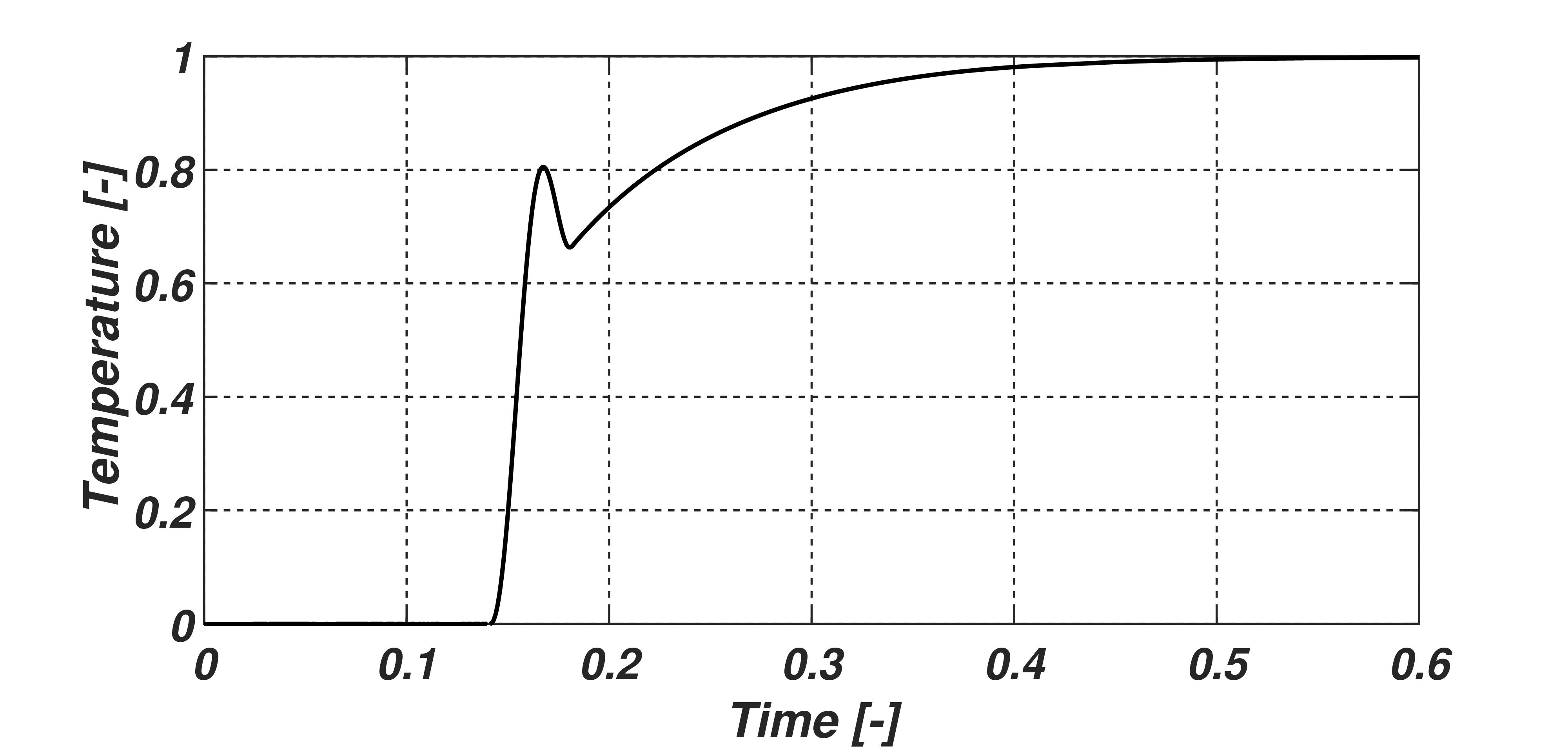}
\caption{The rear-side temperature history related to the MCV equation ($\qtau=0.02$, $\kappa^2=0$), calculated by COMSOL.}
\label{fig:vemcom_mcv}
\end{figure}

Theoretically, it is possible to implant any kind of partial
differential equation within the COMSOL environment. However, to obtain a
solution of a generalized heat equation is not as easy as it seems to
be. Let us begin with the MCV equation. In order to achieve a smooth
solution around the wave front, $100$ elements were used together with
the Runge--Kutta (RK34) time stepping method, which requires $600$ time
steps. Its run time was $44$ s, and for the solution see
Fig.~\ref{fig:vemcom_mcv}.
It is to be noted
that the simulated time interval was shorter, $0.6$ instead of $1$. The
COMSOL solution is
hardly
faster than the explicit scheme presented
above,
and is much slower than the Crank--Nicolson-type implicit scheme. 

\begin{figure}
\centering
\includegraphicss[width=12cm,height=7cm]{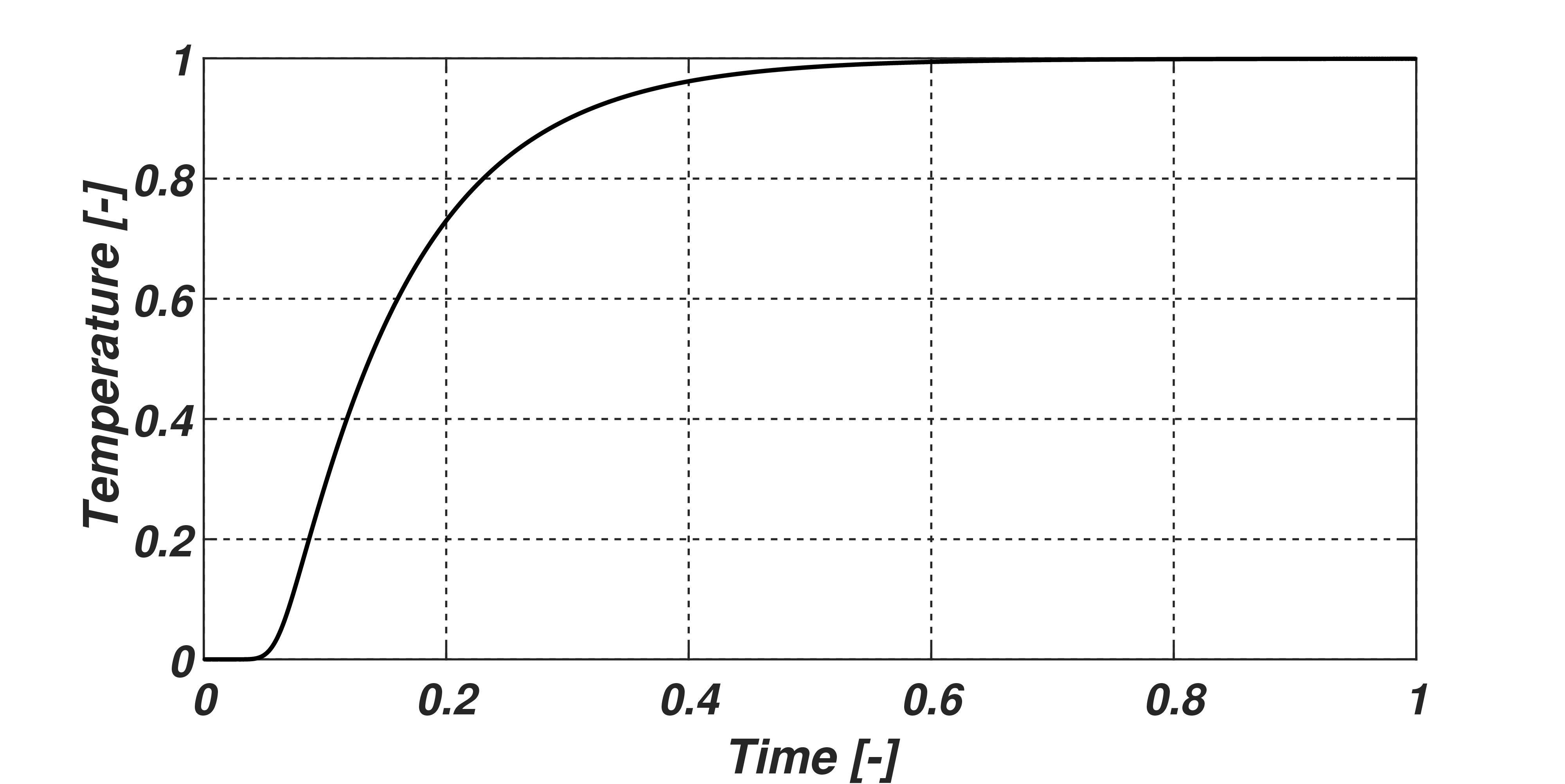}
\caption{The rear-side temperature history related to the GK equation ($\qtau=\kappa^2$), calculated by COMSOL.}
\label{fig:vemcom_gk1}
\end{figure}

\begin{figure}
\centering
\includegraphicss[width=12cm,height=7cm]{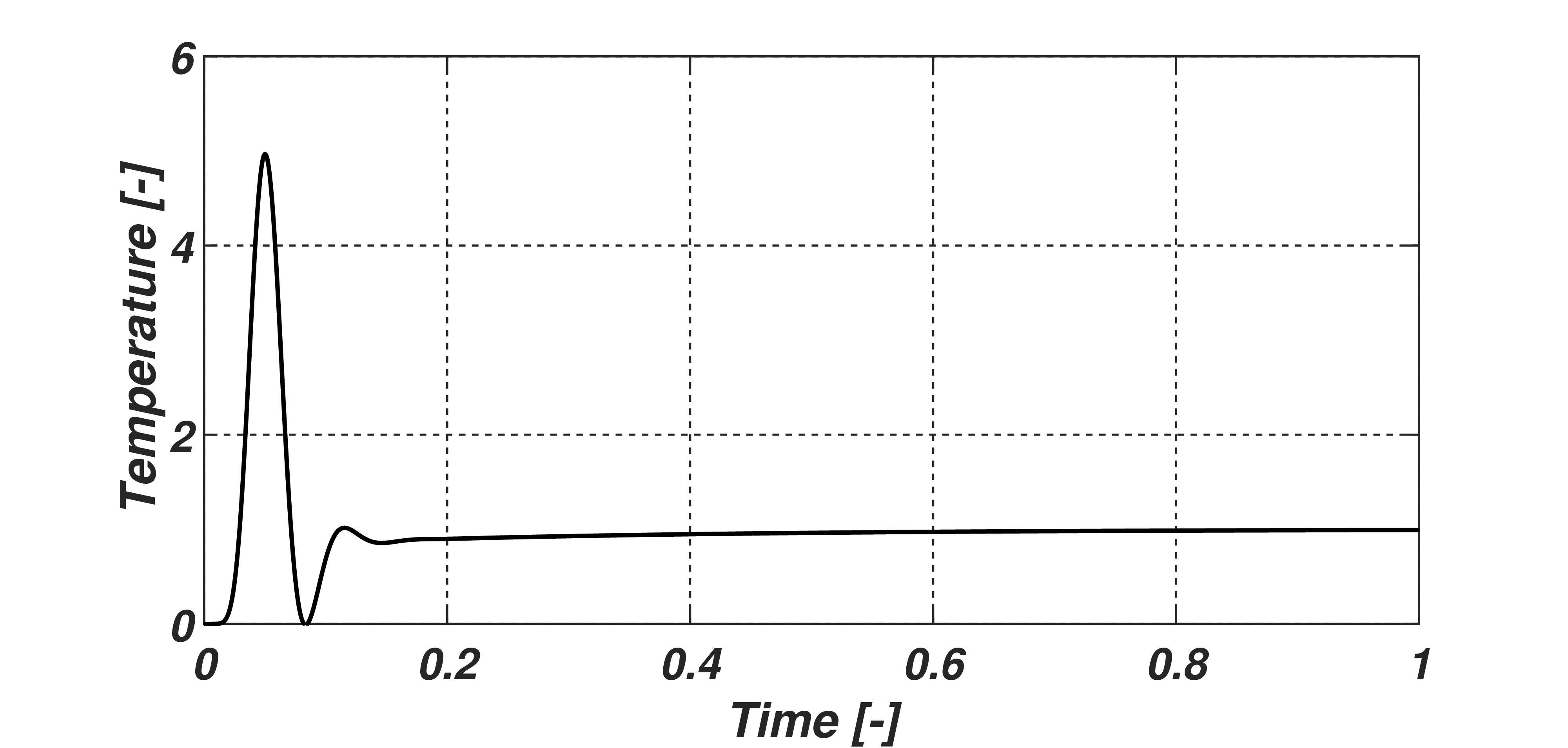}
\caption{The rear-side temperature history related to the GK equation ($\qtau=0.02$, $\kappa^2=0.2$), calculated by COMSOL.}
\label{fig:vemcom_gk2}
\end{figure}

When we turn towards the full GK equation, obtaining the solution is not
straightforward at all. Although COMSOL reproduces
temperature history at the rear side (Fig.~\ref{fig:vemcom_gk1}) in the
Fourier-type special case,
Fig.~\ref{fig:vemcom_gk2} presents a false one with $\qtau=0.02$ and
$\kappa^2=0.2$. Instead of the breakage (like the one in
Fig.~\ref{fig:impgk_anal}), a false wave-shaped solution appears that does
not exist either in the finite difference solution or in the analytical
one.
Moreover, the appearance of this numerical
artifact
 is independent of
mesh and time step sizes, and becomes bigger as $\kappa^2$ is increased. 
Applying again $100$ elements together with a time step of $0.001$, it takes
$65$ s to run (Fig.~\ref{fig:vemcom_gk2}). Hence, COMSOL seems not to be
applicable to solve the GK equation in the highly over-damped domain. 

\section{Outlook for the ballistic-conductive equation}

\begin{figure}
\centering
\includegraphicss[width=12cm,height=7cm]{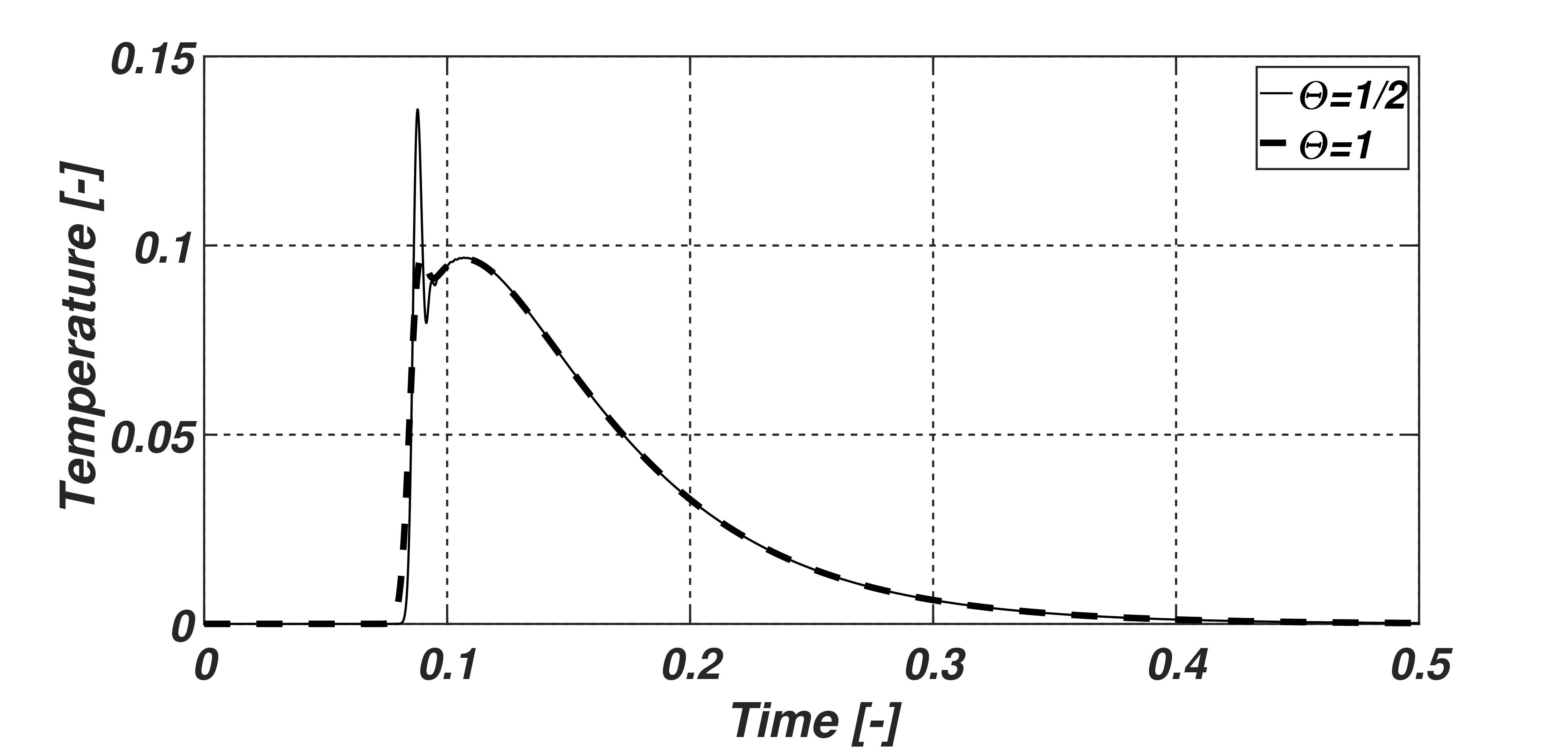}
\caption{The rear-side temperature histories when solving BC equation with schemes $\Theta=1$ and $\Theta=1/2$. The dashed line belongs to $\Theta=1$.}
\label{fig:impbc}
\end{figure}

The ballistic-conductive (BC) equation
is a next-level
generalization beyond the GK equation, and is strongly related to
the low-temperature phenomenon called ballistic heat conduction \cite{KovVan15,
DreStr93a, FriCim95, FriCim96, GrmLeb05}. Indeed, the BC model has been
found to be necessary for explaining low-temperature experiments
\cite{KovVan16, KovVan18}.
Such a hyperbolic equation is more challenging to solve, due to the
double characteristic speed and the sharper jumps in certain parameter
range. It is possible to derive the implicit schemes for this model as
well, as presented below. The
system of partial differential equations in question, in dimensionless
form, reads
 \begin{align}
\qtau \frac{\partial q}{\partial t} +q +\tau_{\Delta}\frac{\partial
T}{\partial x} +\kappa \frac{\partial Q}{\partial x} & = 0,
 \\ \nonumber
\tau_Q \frac{\partial Q}{\partial t} +Q+ \kappa \frac{\partial q}{\partial x}
& = 0,
 \end{align}
and in the discretized form:
 \begin{align}
\frac{\qtau}{\Delta t} (q^{n+1}_j - q^n_j ) + \left ( (1-\Theta) q^n_j +
\Theta q^{n+1}_j \right ) &
 \nonumber \\
+ \frac{\tau_{\Delta}}{\Delta x} \left ( (1-\Theta) ( T^n_j - T^n_{j-1}) + \Theta (T^{n+1}_j - T^{n+1}_{j-1}) \right )
 \nonumber \\
 + \frac{\kappa}{\Delta x} \left ( (1-\Theta) ( Q^n_j - Q^n_{i-1}) + \Theta (Q^{n+1}_j - Q^{n+1}_{j-1}) \right )
& = 0,
 \label{IMPSEPBC} \\ \nonumber
\frac{\tau_Q }{\Delta t} (Q^{n+1}_j - Q^n_j) +\left ( (1-\Theta) Q^n_j + \Theta Q^{n+1}_i \right ) &
 \nonumber \\ \label{BCBC}
+ \frac{\kappa}{\Delta x} \left [ (1-\Theta) (q^n_{j+1} - q^n_j )+ \Theta (q^{n+1}_{j+1} - q^{n+1}_j ) \right ]
& = 0,
 \end{align}
where a further variable $Q$ appears as a current density of
heat flux \cite{KovVan15}. The related relaxation time is denoted by
$\tau_Q$. The system (\ref{IMPSEPBC})--(\ref{BCBC}) can be solved
together with the balance equation of internal energy (\ref{IMPSEPEN}).
Let us use the parameters $\tau_\Delta=0.0076$, $\qtau=0.0186$,
$\tau_Q=0.007$, $\kappa=0.108$, which are taken from \cite{KovVan18} and
are related to the evaluation of a ballistic heat conduction phenomenon.
The same accuracy properties of implicit schemes are experienced as
previously in case of the MCV equation, namely, the one with
$\Theta=1/2$ is more accurate in the vicinity of wave front than the one
with $\Theta=1$, see Fig.~\ref{fig:impbc} for details. It is sufficient
for Crank--Nicolson-type scheme to use $300$ cells with $5000$ time steps
which takes $1.8$ s to solve. 

\begin{figure}
\centering
\includegraphicss[width=12cm,height=7cm]{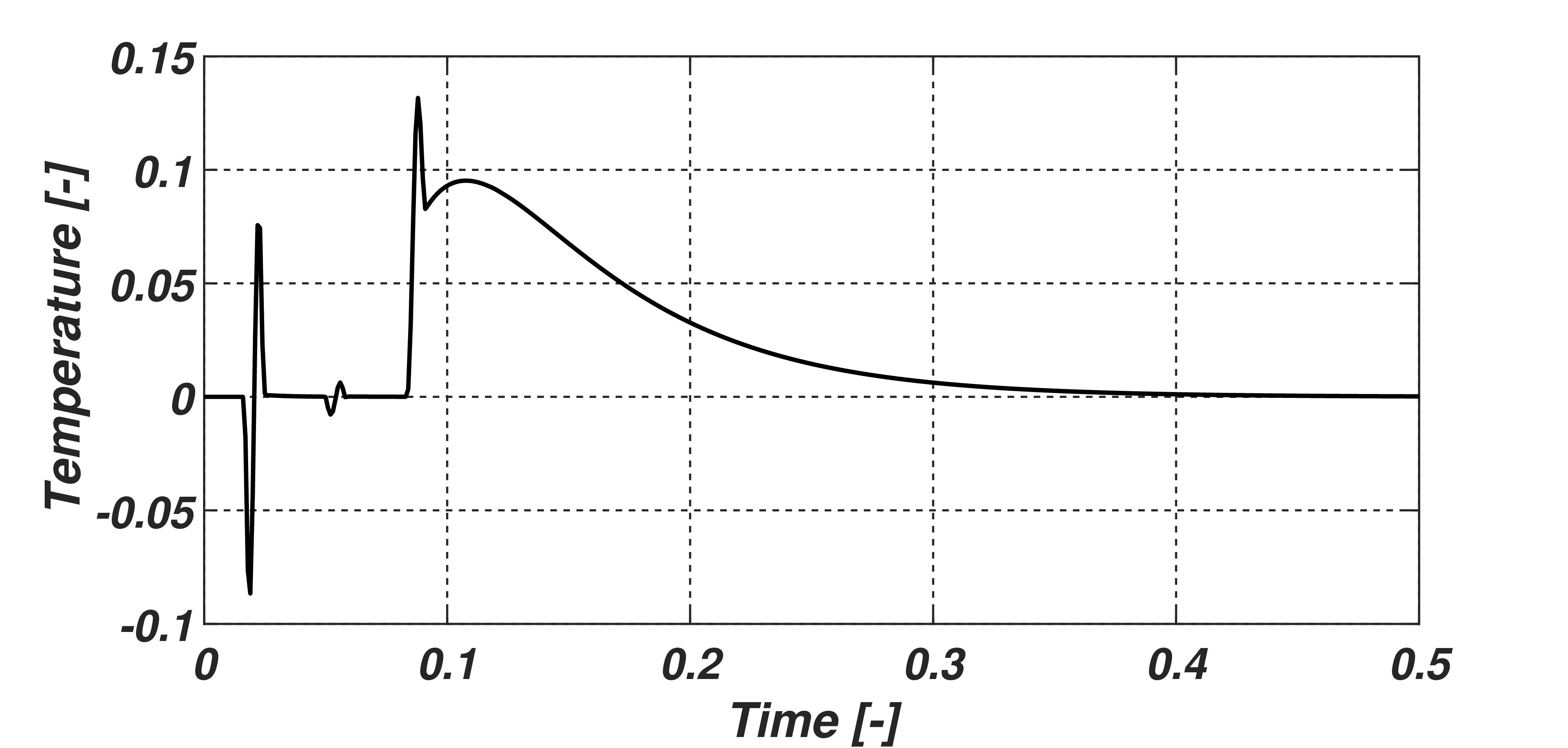}
\caption{The rear-side temperature history related to BC equation, calculated by COMSOL.}
\label{fig:vemcom_bc}
\end{figure}

In contrast, the COMSOL software is much slower and less accurate, i.e.,
it produces the same mesh and time step dependent oscillations and jumps,
see Fig.~\ref{fig:vemcom_bc}.
Only the last jump corresponds to a
real solution, despite of the $1000$ cells used in the simulation with
$9000$ time steps. The run time was $186$ s.
Should one want to avoid these artificial oscillations, the simulation
would require at least ten times more cells and time steps, and it would
take hours for COMSOL to solve the BC model with these settings, in
contrast with the
$1.8$ s run time of the Crank--Nicolson-type approach.

\section{Summary}

Finite difference numerical schemes based on the shifted field concept
have been presented and tested in several cases. It turned out that the
Crank--Nicolson-type implicit scheme is the most accurate, especially in
solving hyperbolic partial differential equations. Not surprisingly, the
presented implicit schemes proved much faster than the explicit one. We
also focused on the validation of numerical schemes using the analytical
solution of the GK equation. It is important to highlight that the
analytical solutions are strongly limited as, for more natural boundary
conditions like heat transfer at the boundary is not yet obtained.
However, having analytical solution is not absolutely necessary in
presence of such a fast and reliable numerical scheme.

The commercial software COMSOL has also been applied for comparison. We
have demonstrated that solving generalized heat equations is challenging
for finite element methods, and leads in some cases to false solutions
so result have to be validated as extensively as possible.

\section{Acknowledgements}
\label{ackn}

This work was supported by the National Research, Development and
Innovation Office of Hungary (NKFIH) via grants NKFIH K116197, K116375,
K124366 and K124508.

\bibliographystyle{elsarticle-num} 




\end{document}